\documentclass[graybox]{svmult}

\usepackage{mathptmx}       
\usepackage{helvet}         
\usepackage{courier}        
\usepackage{type1cm}        
\usepackage{makeidx}         
\usepackage{graphicx}        
\usepackage{multicol}        
\usepackage[bottom]{footmisc}
\usepackage{bm}

\makeindex             

\newcommand\Rey{\mbox{\textit{Re}}}  
\newcommand\Ha{\mbox{\textit{Ha}}}  
\newcommand\Pran{\mbox{\textit{Pr}}} 
\newcommand\N{\mbox{\textit{N}}} 
\newcommand\Gras{\mbox{\textit{Gr}}}  

\begin{document}

\title*{Instabilities in extreme magnetoconvection}
\author{Oleg Zikanov, Yaroslav Listratov, Xuan Zhang, Valentin Sviridov}
\institute{Oleg Zikanov \at University of Michigan - Dearborn, USA, 48128 \email{zikanov@umich.edu}
\and Yaroslav Listratov \at National Research University ``Moscow Power Engineering Institute,'' Moscow,
Russia, 111250 
\and Xuan Zhang \at Max Plank Institute for Dynamics and Self-Organization, G\"ottingen, Germany, 37077 
\and Valentin Sviridov \at Joint Institute for High Temperatures RAS, Moscow, Russia 125412}
%
%
\maketitle

\abstract{Thermal convection in an electrically conducting fluid (for example, a liquid metal) in the presence of a static magnetic field is considered in this chapter. The focus is on the extreme states of the flow, in which both buoyancy and Lorentz forces are very strong. It is argued that the instabilities occurring in such flows are often of unique and counter-intuitive nature due to the action of the magnetic field, which suppresses conventional turbulence and gives preference to two-dimensional instability modes not appearing in more conventional convection systems. Tools of numerical analysis suitable for such flows are discussed.}


\section{Introduction}
\label{sec:intro}
The equivalent terms `magnetoconvection' and `magnetic convection' are used in literature for thermal convection that occurs in electrically conducting liquids in the presence of non-negligible electromagnetic effects \cite{WeissProctor:2014,Ozoe:2005}. The phenomenon  is found in many natural and technological systems. The two well-known examples are the sunspots and planetary dynamos. Their physical mechanisms are not fully understood, but we can with certainty say that they involve thermal convection, Lorentz forces, induction of electric currents and magnetic fields, and, in the case of the sunspots, significant impact of Joule dissipation. 

Less publicized but probably not less important are the occurrences of magnetoconvection in technological systems. In  growth of semiconductor crystals and in casting and remelting of metals, magnetic fields are routinely applied for control of melt flows. The ultimate goal of the control is to optimize the flow state during the solidification so as to achieve higher purity and better properties of the final material at reduced cost and environmental impact \cite{Ozoe:2005,Davidson:2016}. A more recent example is the  liquid metal battery, a promising new device for large-scale stationary energy storage \cite{Kim:2013}. As we have recently learned, its performance is critically influenced by magnetoconvection via enhanced mixing in the liquid metal cathode and electrolyte and deformation of the metal-electrolyte interfaces \cite{Kelley:2014,Shen:2016}.

The area of applications, to which the phenomena presented later in this chapter are particularly relevant, is the nuclear fusion technology. In the currently designed tokamak reactors, liquid-metal components are considered  as parts of a promising technical solution for the blanket, divertor and first wall that form the inner part of the toroidal vessel containing the plasma \cite{Dolan:2013,Abdou:2015}. A Li-containing metal (most likely PbLi alloy) flowing within the blanket and the divertor will absorb the highly energetic neutrons generated in the fusion reaction thus achieving the three-fold goal of: (a) converting the neutron's energy into heat and transporting the heat into an external power-generation cycle, (b) shielding the reactor's exterior, and (c) breading the tritium fuel for the reaction. For the blanket and divertor, the technology will be tested, for the first time in realistic conditions, in the ITER experimental reactor, and, hopefully, developed further in the next-generation DEMO and FNSF facilities \cite{Abdou:2015}. 

The use of liquid metals at the first wall (the wall facing the plasma) aims at the critical and yet unsolved problem of wall damage by the radiation heat flux. Apart from the evident danger to the reactor's structure, the damage, even a minor one, would lead to contamination of plasma. An interesting idea of resolving this issue is that of a `self-healing' cover of the wall formed by flowing liquid lithium. Several variously realistic concepts exist, including an entirely liquid film stabilized by the magnetic field \cite{apex:1999}, a porous matrix saturated by Li \cite{Evtikhin:2000}, and a wall made of tiles with narrow Li-filled open channels \cite{Ruzic:2011}.

The long-understood potential of liquid-metal components for the fusion technology has resulted in serious efforts toward understanding of liquid metal flows in very strong (4 to 12 T is expected in the ITER and DEMO reactors) static magnetic fields. The picture of the rather peculiar flow behavior that has emerged from these studies includes suppression of turbulence and strong magnetohydrodynamic resistance in flows through ducts with electrically conducting walls (see section \ref{sec:trans} of this chapter for  further discussion). Unfortunately, in hindsight, the picture appears to have little relevance to the real flows in the reactor's components. While  correct for isothermal flows, it fails to take into account the effect of the thermal convection caused by the very strong heat flux imposed on the fluid by the radiation from plasma and by the absorbed neutrons. The cumulative steady heat flux varies, depending  on the reactor type and the location within the system, between about 4 and 20 MW/m$^2$. Much higher pulse load is expected during the plasma disruption events.

We will see in the following discussion that under the simultaneous impact of very strong thermal convection and very strong magnetic field,  the nature of the flows changes in the way that is drastic, paradoxical and cannot be predicted on the basis of understanding of the effects of convection and magnetic field taken separately. This leads to profound changes in the system's operation. 

The term `extreme magnetoconvection' will be used for such situations. While originating in the studies of configurations directly relevant to the nuclear fusion technology, the concept is universal. It applies to every situation, in which a flow  of an electrically conducting fluid is affected simultaneously by strong heating and strong magnetic field. 

\section{Magnetohydrodynamic model and governing equations}
\label{sec:model}
Our discussion will focus on flows of incompressible, single-phase, electrically conducting but not magnetizable fluids, such as, for example, liquid metals. The fluids have high electric conductivities, so the effects of electromagnetic fields are strong. The physical properties of the fluids are assumed constant with exception of the density in the buoyancy force term, which is treated using the Boussinesq approximation.

The quasi-static (low magnetic Reynolds number) approximation of the electromagnetic effects is used. It has been derived theoretically and validated in comparison with experiments (see, e.g. \cite{Davidson:2016} for a discussion). The approximation is valid when the magnetic Prandtl and Reynolds numbers are small: 
\begin{equation}
\label{qust}
\Pran_m\equiv \sigma \mu_0 \nu \ll 1, \: \Rey_m\equiv UL \sigma \mu_0=\Rey\Pran_m \ll 1,
\end{equation}
where $U$ and $L$ are the relevant velocity and length scales, $\sigma$ and $\nu$  are the electric conductivity and kinematic viscosity of the fluid and and $\mu_0=4\pi\times 10^{-7}$ H/m is the magnetic permeability of vacuum, and $\Rey$ is the hydrodynamic Reynolds number defined below. Under such conditions, the perturbations of the magnetic field induced by velocity fluctuations are negligibly weak in comparison with the imposed field and adjust nearly instantaneously to the change of velocity. The two-way coupling between the flow and the magnetic field is, with good accuracy, reduced to the one-way effect of the field on the flow. 

The non-dimensional governing equations can be written, in a general form, as:
\begin{eqnarray}
\hskip-15mm& & \frac{\partial \bm{u}}{\partial t} +
(\bm{u}\cdot \nabla)\bm{u} = -\nabla p 
+  \frac{1}{\Rey}\nabla^2 \bm{u} + \bm{F}_b+\bm{F}_L,\label{nsviscous}\\
\hskip-15mm& & \nabla \cdot\bm{u}  =  0,\label{incompr}\\
\label{scalar} \hskip-15mm& & \frac{\partial T}{\partial t} +
\bm{u}\cdot \nabla T = \frac{1}{\Rey\Pran} \nabla^2 T + q.
\end{eqnarray}
where $\bm{u}$, $p$ and $T$ are the fields of velocity, pressure, and temperature  perturbation with respect to an appropriate reference value, and we permit existence of internal heating of volumetric power $q$ (e.g. by absorbed neutrons in a
fusion reactor blanket). The buoyancy and Lorentz forces are:
\begin{eqnarray}
\bm{F}_b & = &  -\Gras \Rey^{-2} T\bm{e}_g,\label{fbu}\\
\bm{F}_L & = &  \Ha^2\Rey^{-1} \bm{j}\times \bm{B},\label{flo}
\end{eqnarray}
where $\bm{e}_g$ is the unit vector in the direction of gravity and $\bm{B}$ is the non-dimensional imposed magnetic field. The electric current density $\bm{j}$ is given by the Ohm's law 
\begin{equation}
\label{ohm}
\bm{j}=-\nabla \phi + \bm{u} \times \bm{B},
\end{equation}
where the electric potential $\phi$ is a solution of the Poisson equation expressing the instantaneous electric neutrality of the fluid:
\begin{equation}\label{potent}
\nabla^2\phi = \nabla \cdot (\bm{u}\times \bm{B}).
\end{equation}

The equations are non-dimensionalized using some typical scales $L$, $U$, $\Delta T$, and $B_0$ of length, velocity, temperature variation, and the magnetic field strength, and the derivative scales $\rho U^2$, $\sigma U B_0$, and $LUB_0$ for pressure, electric current density, and electric potential.

The non-dimensional parameters are the Reynolds number
\begin{equation}\label{reynolds}
\Rey\equiv \frac{UL}{\nu},
\end{equation}
the Prandtl number
\begin{equation}\label{prandtl}
\Pran\equiv\frac{\nu}{\chi},
\end{equation}
which is typically in the range between 0.01 and 0.1 for practically interesting fluids,
the Hartmann number
\begin{equation}\label{hartmann}
\Ha\equiv  B_0L\left(\frac{\sigma}{\rho\nu}\right)^{1/2},
\end{equation}
and the Grashof number
\begin{equation}\label{grashof}
\Gras\equiv  \frac{g\beta \Delta T L^3}{\nu^2},
\end{equation}
where $\chi$ and $\beta$ are the temperature diffusivity and expansivity  of the liquid.

The square of the Hartmann number is an estimate of the magnitude of the ratio between the Lorentz and viscous forces. As an alternative, the Stuart number (the magnetic interaction parameter) 
\begin{equation}
\label{stuart}
\N\equiv \frac{\Ha^2}{\Rey}=\frac{\sigma B_0^2L}{\rho U}
\end{equation}
estimating the ratio between the Lorentz and inertial forces can be used.

The typical boundary conditions at solid walls include the no-slip conditions for velocity, prescribed temperature or prescribed heat flux, and zero normal current or prescribed electric potential. In some configurations, the finite electrical or thermal conductivity of the wall plays an essential role, so a conjugate transport problem in three-dimensional form or in the framework of a thin-wall approximation must be solved for $T$ or $\phi$. In the examples presented in section \ref{sec:results} of this chapter, the walls are assumed to be electrically perfectly insulated, so the condition of zero normal gradient of $\phi$ is applied. For flows in pipes and ducts, the typically used inlet and exit conditions are not different from those used for  flows without electromagnetic effects.

A brief discussion is in order concerning the accuracy of the approximations used in our model in the case of real systems with magnetoconvection. The magnetic Prandtl number is typically very small ($\sim 10^{-5}$), so the conditions (\ref{qust}) are satisfied by practically all laboratory and industrial flows of liquid metals. It is generally believed that the quasi-static model is acceptably accurate for such flows, although second-order non-quasi-static effects are possible and still need to be investigated. For geo- and astrophysical systems, $\Rey_m$ based on the system size and integral velocity is not small, so the quasi-static model must, in principle, be replaced by the full MHD model that includes the perturbations of the magnetic field. At the same time, $\Pran_m$ remains small, so the physical phenomena found in a quasi-static analysis remain qualitatively relevant for processes at smaller length scales.

The  approximation, in which all the physical properties are assumed constant except density, for which the linear Boussinesq model is used, may also seem an oversimplification. As we discuss below, the magnetoconvection flows often exhibit very strong variations of temperature. The analysis of the available data, however, shows that the variation of properties with temperature is rather weak for the relevant fluids. For example, for the PbLi alloy in the temperature range typical for fusion reactor applications, only dynamic viscosity and thermal conductivity vary significantly \cite{MasdelesValls:2008}. At $T=600^{\circ}C$, they change by, respectively, about 10\% and 5\% when temperature changes by 50 K.

\section{Flow transformation in the presence of a strong magnetic field}
\label{sec:trans}
The  modification of a flow of an electrically conducting fluid by an imposed magnetic field has, on the level of basic physics, three main aspects, all related to the induction of electric currents in a moving fluid (see, e.g. \cite{Davidson:2016,Mueller:2001}). Firstly, the Lorentz force (\ref{flo}) changes the global structure of the flow. Two major manifestations of the change are the formation of MHD boundary layers near solid walls and the strong magnetohydrodynamic pressure drop in flows through ducts, especially in the case of electrically conducting walls. Secondly, fluctuations of velocity are suppressed. This is caused by the Joule dissipation of the fluctuation-induced electric currents, which can be seen as an additional mechanism of conversion of flow's kinetic energy into heat. The third, subtler aspect is related to the fact that the local rate of the Joule dissipation is proportional to the square of the gradient of flow velocity along the field lines \cite{Moffatt:1967,Sommeria:Moreau:1982}. Flow structures with largest such gradients experience the strongest suppression, while the structures uniform along the field lines are not affected. The flow becomes anisotropic, with its structures elongated in the direction of the field (examples can be found, e.g. in \cite{Zikanov:1998,Vorobev:2005,Thess:2007,Krasnov:2012}).

The modification becomes significant at $\N\sim 1$. At substantially larger $\N$, it manifests itself as:
\begin{enumerate}
\item Full suppression of three-dimensional turbulence,
\item Profound modification of the flow's global structure and formation of very thin MHD boundary layers,
\item Quasi-two-dimensionality, with the flow structures uniform along the field lines outside the boundary
layers (purely 2D flows are found in absence of walls non-parallel to the magnetic field \cite{Zikanov:1998,Thess:2007,Zhao:2014,Dong:2012}).
\end{enumerate}

Our interest is on the situations when the convection itself and the effect of its modification by the magnetic field are both very strong. This can be characterized as:
\begin{equation}
\label{situation}
\Gras\gg 1, \: \Ha\gg 1, \: \N>1.
\end{equation}
At first glance, the commonly accepted picture of flow transformation discussed above implies that turbulence is fully suppressed in such cases and, therefore, the
flows should all be steady and laminar. This should be true at arbitrarily large $\Gras$ and $\Rey$, provided the magnetic field is sufficiently strong. This view was, in fact, for long time dominant in theoretical studies and applied research, e.g. in the areas of crystal growth, metallurgy, and nuclear fusion technology. The recent computations and experiments, however, proved the view to be incorrect. 

The key was the better understanding of the flow instabilities (caused by convection, shear, or some other mechanism) to perturbations with zero or weak gradients along the field lines. Such perturbations are not suppressed or suppressed only weakly even at large values of $\Ha$ and $\N$. At large $\Gras$ and $\Rey$, their unbridled growth leads to flow structures, which, while not turbulent in conventional sense, are strong, spatially complex and possibly unsteady. Their nonlinear dynamics may result in unusual flow states, such as two-dimensional turbulence, high-amplitude quasi-periodic fluctuations or intermittency in the form of violent bursts of three-dimensional unsteady behavior interrupting long laminar periods.  Two interesting earlier examples of the intermittency can be found in the simulations of the channel flow with spanwise magnetic field \cite{Boeck:2008} and of the periodic-box flow driven by artificial large-scale forcing \cite{Zikanov:1998}. Further examples are discussed below.


\section{Extreme magnetoconvection}\label{sec:results}
The phenomenon of extreme magnetoconvection is explained in this section using several examples taken from recent studies. The work is originally motivated by the questions of the fusion technology, although, as we will see, the phenomenon itself is of general nature and has to be expected in every flow satisfying the conditions (\ref{situation}). 

The first indications that turbulence suppression does not necessarily lead to a steady laminar convection were obtained in the experiments, such as \cite{Genin:2011Pamir,Melnikov:2016,Kirillov:2016}. Mixed convection in flows of mercury in pipes and ducts of various orientations with respect to gravity and imposed magnetic field were studied at $\Gras\le 10^8$, $\Ha\le 800$ $\Rey\le 5\times 10^4$. Walls of the test section were partially heated and partially thermally insulated. It has been found that conventional turbulent fluctuations of velocity and temperature are always suppressed by sufficiently strong magnetic fields. At even stronger fields, however, the so-called anomalous fluctuations appear. This  flow regime is not turbulent, but is characterized by nearly regular (with a few dominating frequencies), low-frequency, high-amplitude oscillations of temperature and velocity.

\begin{figure}
\begin{center}
\includegraphics[width=0.65\textwidth]{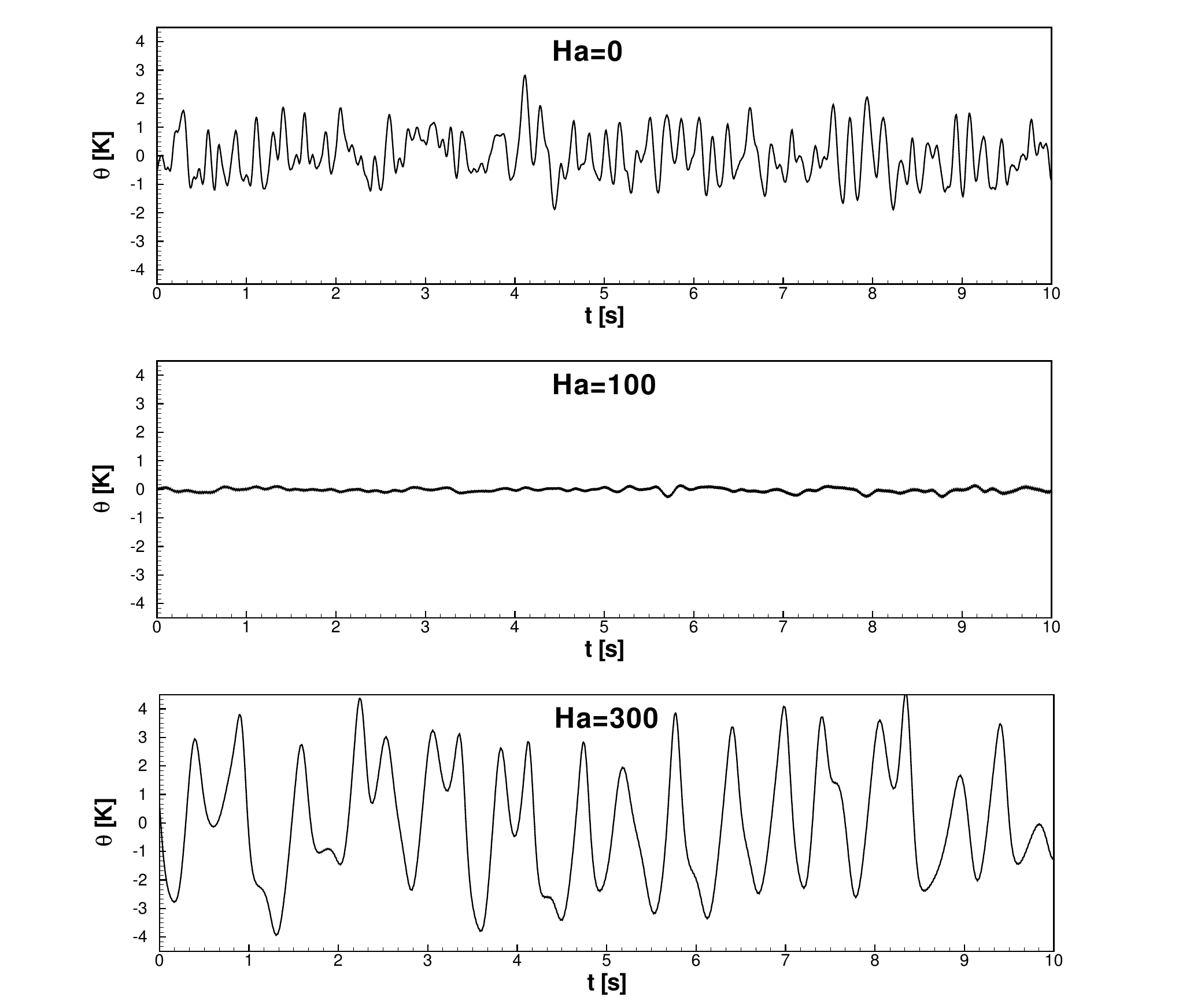}\\
\caption{Mixed convection in a horizontal pipe with lower half of the wall heated and transverse horizontal magnetic field. $\Gras=8.3\times 10^7$, $\Rey=10^4$, $\Ha=0$, 100, 300.  Times signals of temperature measured in \cite{Genin:2011Pamir} are shown. }
\label{fig1}
\end{center}
\end{figure}

As the first illustration of this regime, Fig.~\ref{fig1} shows the results of the experiments \cite{Genin:2011Pamir}  for a flow in a horizontal pipe with the lower half of the wall heated, the upper half thermally insulated, and the magnetic field imposed in the transverse horizontal direction. We see that at a moderately strong magnetic field ($\Ha=100$ in the figure),  turbulence is suppressed resulting in a steady-state signal of temperature. At stronger magnetic fields ($\Ha=300$ shown in the figure and higher $\Ha$, not shown but found in the experiment), however, fluctuations reappear, but now in the form of quasi-regular high-amplitude oscillations. 

\begin{figure}
\begin{center}
\includegraphics[width=0.49\textwidth]{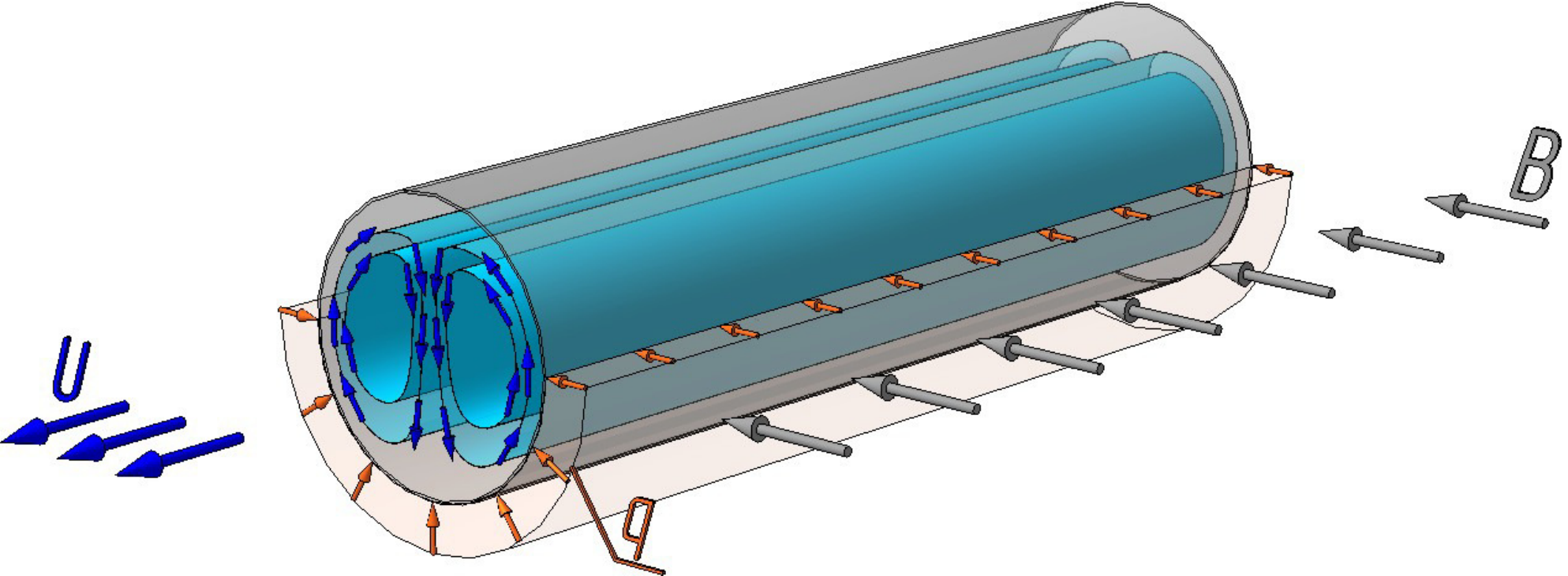} \includegraphics[width=0.49\textwidth]{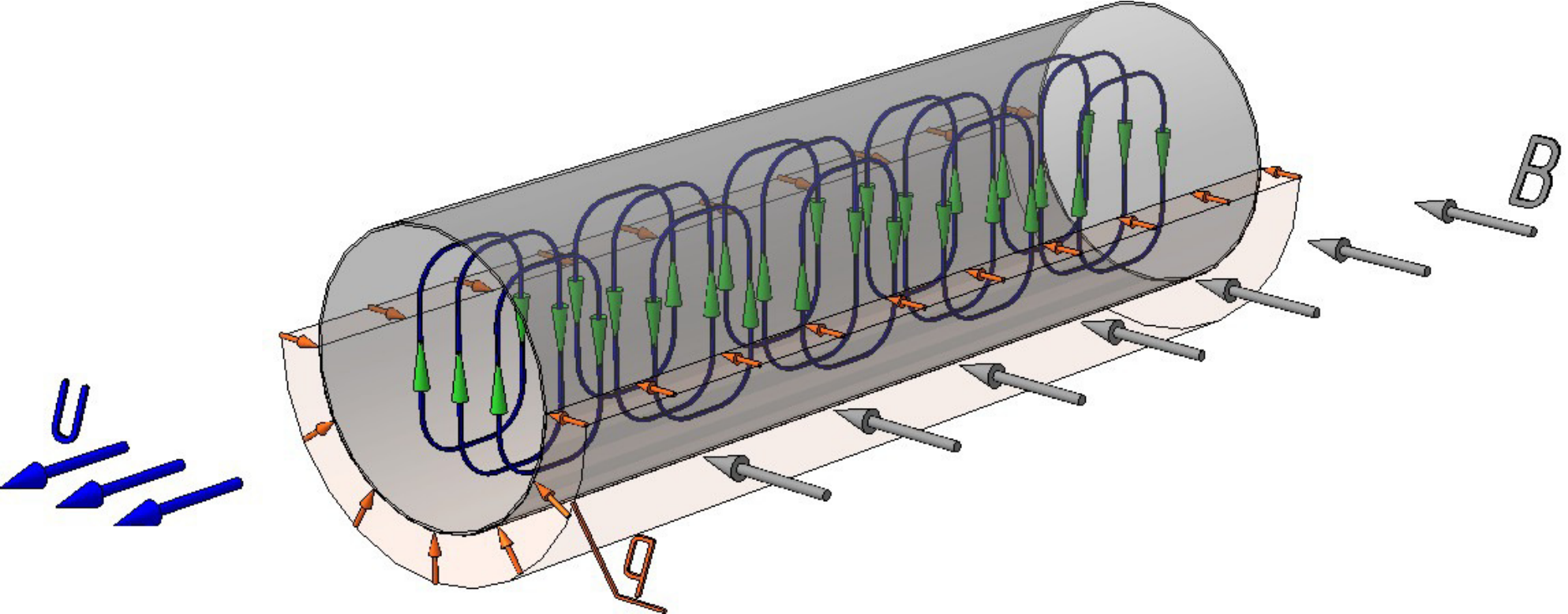}\\
\vspace{5mm}
\includegraphics[width=0.75\textwidth]{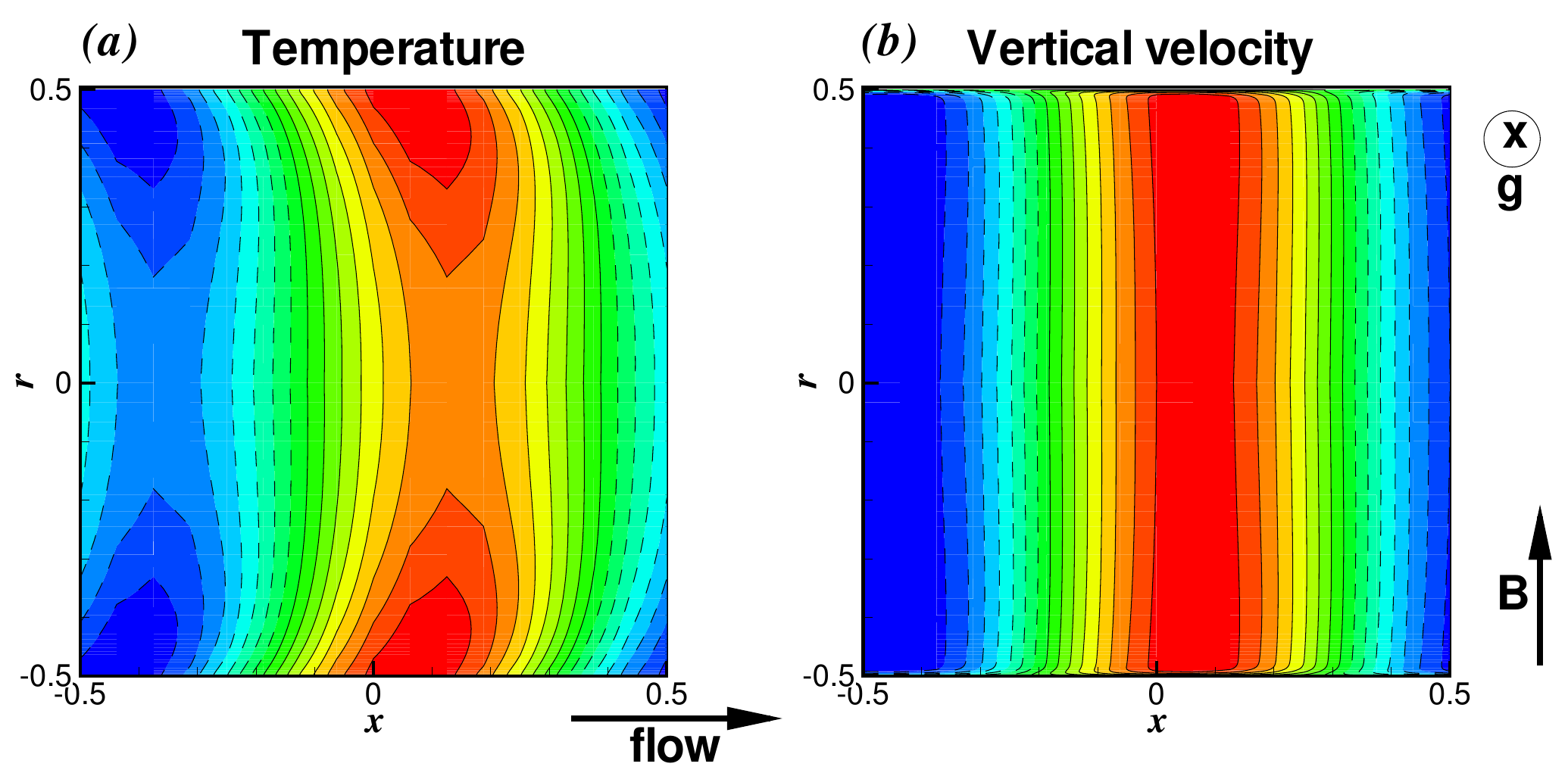}\\
\caption{Mixed convection in a horizontal pipe with lower half of the wall heated and transverse horizontal magnetic field. $\bm{B}$, $\bm{q}$, $\bm{U}$ and $\bm{g}$ indicate the directions of the imposed magnetic field, wall heat flux,  mean axial velocity and gravity acceleration, respectively. Top row:  The schematic representation of convection  structures in the presence of moderately strong (left) and very strong (right) magnetic field. Bottom row: Results of the linear instability modeling \cite{ZikanovJFM:2013}. The most unstable mode at $\Gras=8.3\times 10^7$, $\Rey=10^4$, $\Ha=300$, and axial wavenumber $\lambda=1.0$ is shown using the distributions of temperature and vertical velocity perturbations in the horizontal cross-section through the pipe's axis. Solid and dashed contour lines correspond, respectively, to positive and negative values.}
\label{fig2}
\end{center}
\end{figure}

The nature of the flow transformation was revealed in the computational analysis \cite{ZikanovJFM:2013}, which included modeling of  linear stability  and  high-resolution three-dimensional simulations of the flows in  the entire test section of the experiment. It has been found that at moderate $\Ha$ the laminar convection flow has the structure of two steady streamwise rolls superimposed on the axial flow along the pipe (see the diagram in the upper left plot of Fig.~\ref{fig2}). At stronger magnetic fields, the rolls are suppressed due to their strong gradients along the field lines.  This opens the way for the growth of convection instability modes in the form of transverse rolls (see the upper right diagram in Fig.~\ref{fig2} and the numerical results in the bottom row of Fig.~\ref{fig2}), which have axes along the magnetic field and, therefore, experience much weaker suppression (only substantial in the near-wall regions). The rolls grow to high amplitudes and, as they are transported along the pipe by the mean flow, cause the fluctuations of temperature seen in the experiment. The explanation is supported by the good quantitative agreement between the simulations and the experiment in terms of the typical amplitude and main frequency of the temperature oscillations. We also note that 
a similar flow transformation is found in the simulations of mixed convection in a horizontal duct with bottom heating \cite{Zhang:2014}.

\begin{figure}
\begin{center}
\includegraphics[width=0.62\textwidth]{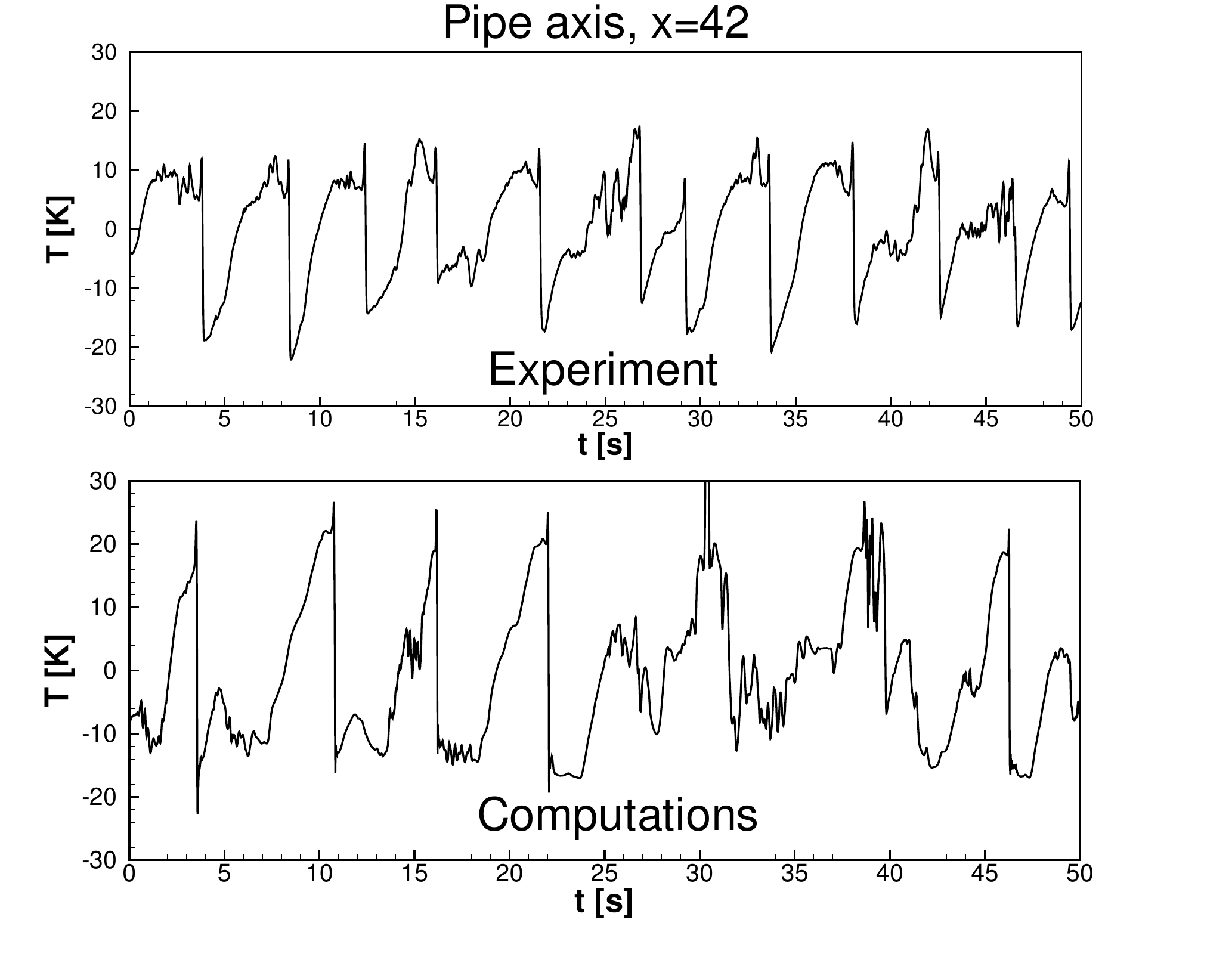}\includegraphics[width=0.37\textwidth]{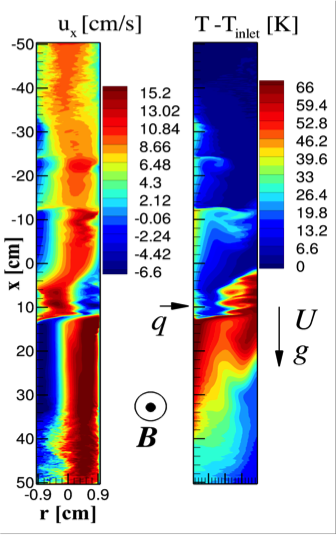}\\
\caption{Mixed convection in a vertical pipe with downward mean flow at $\Rey=1.2\times 10^4$, $\Gras=1.2\times 10^8$, $\Ha=300$. In the test section of the pipe (at $-41<x<41$ cm) half of the wall is heated with the constant rate $q$, the other half is thermally insulated, and the  horizontal magnetic field $\bm{B}$ directed perpendicularly to the temperature gradient is imposed. Left column: Comparison between experimentally observed \cite{Melnikov:2016} and computed \cite{Zikanov:2016} temperature fluctuations at $r = 0$, $x = 21$ cm. Right column: Results of the DNS \cite{Zikanov:2016} of the experiment. Instantaneous distributions of computed streamwise velocity and temperature in the axial cross-section normal to the magnetic field are shown in the dimensional units of the experiment.}
\label{fig4}
\end{center}
\end{figure}

An even more striking example of the anomalous temperature fluctuations is observed in vertically oriented pipes and ducts \cite{Melnikov:2016} (see Fig.~\ref{fig4}). At large $\Ha$ and $\Gras$, slow oscillations of temperature of very high amplitude are detected. Before discussing their nature we need to mention the so-called elevator modes, which are the exact solutions of the Boussinesq equations in vertically unbounded systems with unstable stratification (see, e.g. \cite{Batchelor:1993,Calzavarini:2006,Schmidt:2012}). They have the form of ascending/descending jets exponentially growing in time. In hydrodynamic systems, the jets are typically no more than a mathematical artifact, since their growth rapidly causes  three-dimensional instabilities and breakdown into turbulence. 

The situation is different in the magnetoconvection systems, where the three-dimensional instabilities can be suppressed by the magnetic field, so the jets can grow to large amplitudes before succumbing to large-scale instabilities to quasi-two-dimensional perturbations with weak gradients along the field lines. The best studied system is the downward flow in a vertical pipe or duct with heating applied to a part of or entire wall (see Fig.~\ref{fig4}). The unstable stratification, with the mean temperature growing linearly downward, is formed in this flow as a result of energy balance between the wall heating and axial convective heat transfer. The numerical analysis of such a flow with  transverse magnetic field  \cite{Liu:2015} have shown that in a vertically unlimited duct, the elevator modes are stabilized and become a dominant component of the solution at large $\Gras$. 

The more complex situation of realistic vertically bounded flow domain was analyzed in the recent numerical simulations \cite{Zikanov:2016}. The DNS of the experimental setup of \cite{Melnikov:2016} reproduced the temperature oscillations and demonstrated that they were caused by the axially finite but elongated ascending and descending jets  (see Fig.~\ref{fig4}). The jets are stabilized by the magnetic field, so they grow to large amplitudes before eventual development of shear instability to perturbations nearly uniform along the magnetic field. The instability leads to  quasi-periodic breakdowns and formation of localized zones of strong mixing. Transport of the zones by the mean flow creates the cycles of growth and sharp drop of local temperature observed in the experiments.

Our final example is the magnetoconvection in a long horizontal duct with axial magnetic field. The configuration was a subject of recent computational studies \cite{Zhang:2015,Zhang:2017,Zhang_P1:2017,Zhang_P2:2017}. It has practical relevance to the fusion reactor technology, where it can be considered as a model of the flow in a long toroidally (horizontally and along the main component of the magnetic field) oriented duct within the blanket.

If we assume that  the duct is long and ignore the end effect, for example using periodic inlet-exit conditions, the system permits purely two-dimensional (axially uniform) solutions. Such solutions are expected in the presence of a sufficiently strong axial magnetic field. In principle, they should be considered as approximations, since in an infinite domain and for an arbitrarily strong magnetic field one can introduce perturbations of the wavelength sufficiently large to survive the magnetic damping  (see \cite{Thess:2007} for a discussion). Practically, however, the analysis of \cite{Zhang:2015} shows that at the typical parameters of a fusion reactor, the axially non-uniform component of the velocity is very weak, and the flow can be approximated as two-dimensional with good accuracy.

\begin{figure}
\begin{center}
\includegraphics[width=0.9\textwidth]{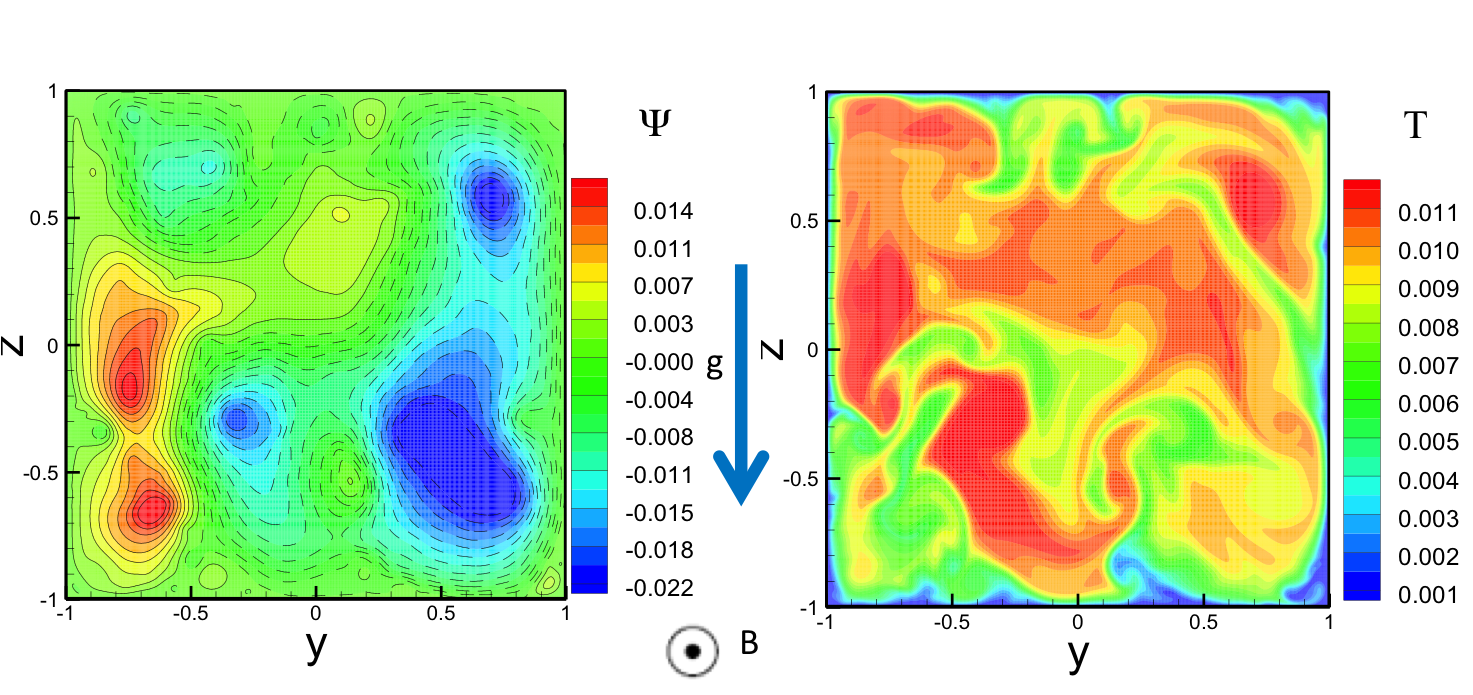}\\
\caption{Flow in a horizontal duct with zero mean velocity and thermally perfectly conducting walls \cite{Zhang:2015}. The fluid is heated internally with the volumetric heating rate decreasing exponentially with the distance to the wall at $y=-1$. The flow is assumed two-dimensional ($x$-independent) due to the presence of very strong axial magnetic field. Instantaneous distributions of streamfunction (plot on the left) and temperature (plot on the right) are shown for $\Gras=10^{11}$.}
\label{fig5}
\end{center}
\end{figure}

Convection caused by non-uniform internal heating (the volumetric heating rate decreasing exponentially with the distance to a sidewall) was analyzed in  \cite{Zhang:2015,Zhang_P1:2017,Zhang_P2:2017}. It was found in \cite{Zhang:2015} that at zero mean flow and thermally perfectly conducting walls, the system demonstrates classical two-dimensional turbulence with strong mixing and the Nusselt number growing as $\sim \Gras^{0.2}$ (see Fig.~\ref{fig5}). 

\begin{figure}
\begin{center}
\includegraphics[width=0.9\textwidth]{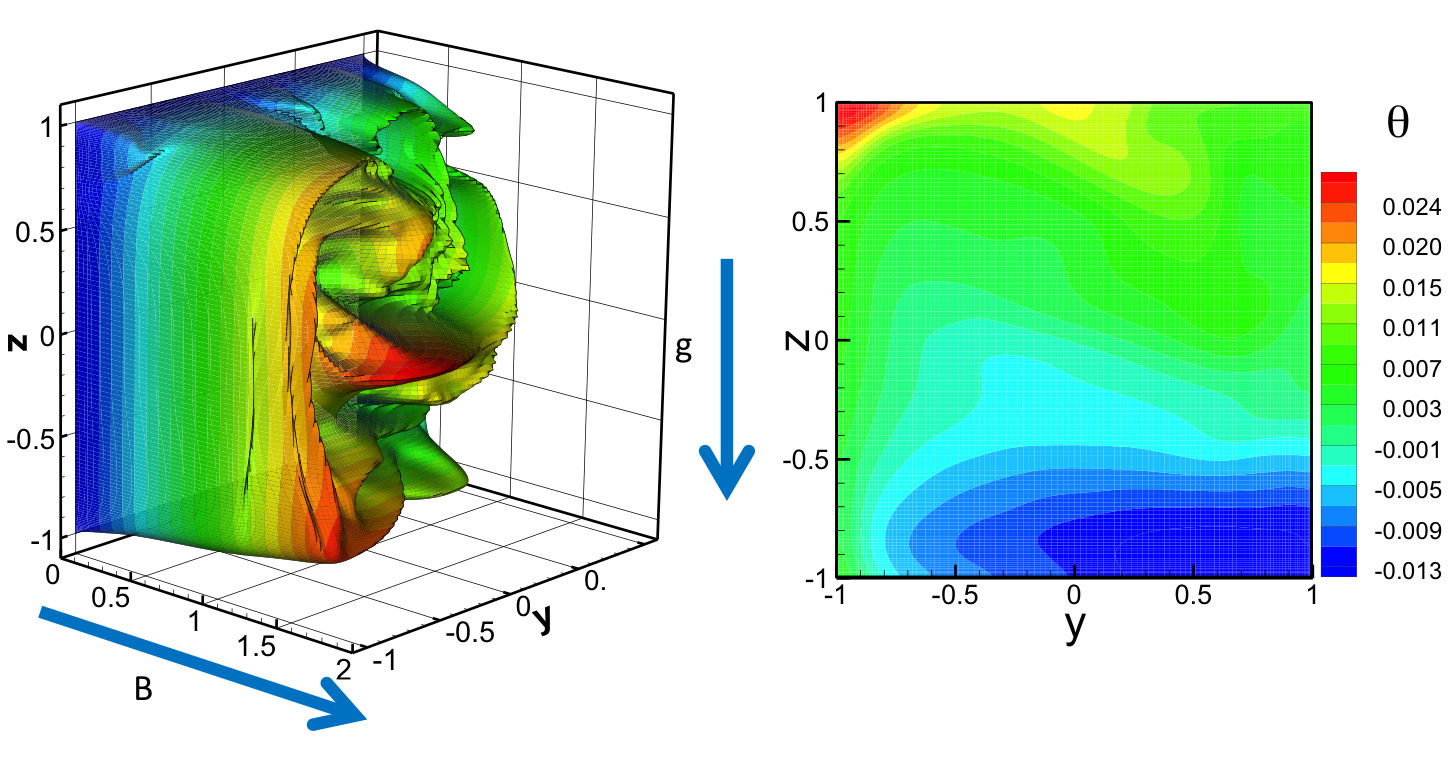}\\\includegraphics[width=0.9\textwidth]{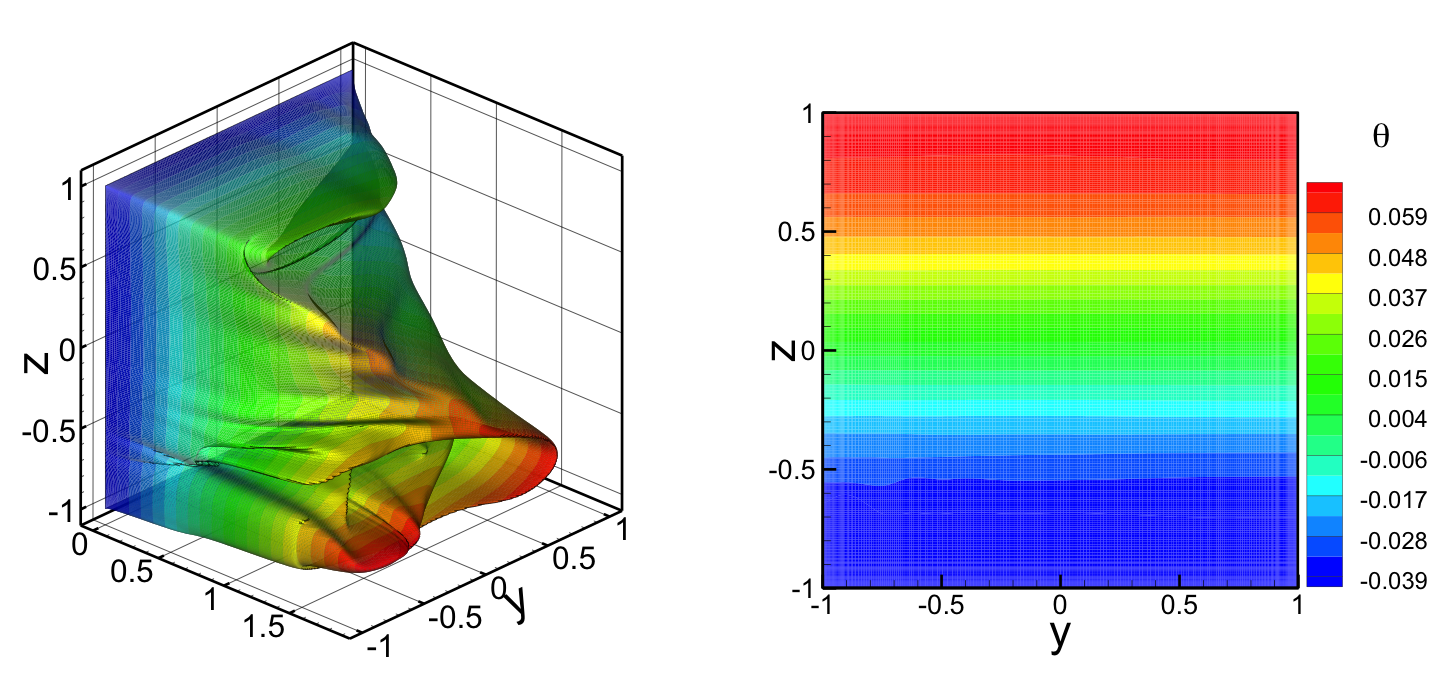}\\
\caption{Flow in a horizontal duct with non-zero mean velocity and thermally perfectly insulating walls \cite{Zhang_P2:2017}. The fluid is heated internally with the volumetric heating rate decreasing exponentially with the distance to the wall at $y=-1$.. The flow is assumed two-dimensional ($x$-independent) due to the presence of very strong axial magnetic field. Instantaneous distributions of axial velocity (left column) and temperature (right column) are shown for $\Rey=10^6$, $\Gras=10^{9}$ (top row) and $\Rey=5\times 10^5$, $\Gras=10^{11}$ (bottom row).}
\label{fig6}
\end{center}
\end{figure}

Further studies have shown that the two-dimensional turbulence regime is highly sensitive to the features of the blanket system not taken into account in the simple model of \cite{Zhang:2015}. One such feature is the vertical (poloidal) component of the magnetic field inevitably present in the blanket. While much weaker than the main toroidal component, it still produces sizable magnetic damping effect. As shown in \cite{Zhang_P1:2017}, it suppresses turbulence and leads to the convection pattern dominated by large-scale two-dimensional rolls. Interestingly, the thin shear layers between the rolls are susceptible to local Kelvin-Helmholtz instabilities, so the entire flow retains three-dimensionality at higher $\Ha$ than in the case of zero poloidal magnetic field \cite{Zhang:2017}.  

A peculiar  behavior was discovered in \cite{Zhang_P2:2017}, where the flow in a horizontal duct with thermally insulated walls was considered. The energy deposited by the internal heating was diverted by the mean flow along the duct. As illustrated in Fig.~\ref{fig6}, noticeably strong convection flow within the transverse cross-section of the duct develops only at moderate values of $\Gras$. At large $\Gras$ (approximately at $10^9$ and larger), the convection is suppressed by the stable stratification of temperature. The mechanism leading to this effect is explained in \cite{Zhang_P2:2017}. Briefly, the origin is the distribution of the mean temperature, which increases linearly downstream. The associated buoyancy force causes redistribution of the axial gradient of pressure in the transverse cross-section, which in turn leads to top-bottom asymmetry of the axial velocity (see Fig.~\ref{fig6}) and the associated axial convective heat transfer.

The examples described in this section illustrate the meaning of the term `extreme magnetoconvection'. It applies to flows at very large $\Gras$, in which we would normally expect turbulence and well-mixed temperature field. Three-dimensional conventional turbulence is, however, fully suppressed by a strong magnetic field. As the mixing effect is removed, the temperature gradients grow to unusually high values. The flow eventually becomes unstable, but to the perturbations of the special kind least affected by the magnetic field: perfectly or nearly two dimensional (uniform along the field lines) and, therefore, large-scale. This often results in unexpected and extreme flow behaviors characterized by high-amplitude, slow, quasi-regular fluctuations, thin jets, cold and hot spots, two-dimensional turbulence, etc. The initial mechanism of the instability is always thermal convection. The shear layer instability also often plays a role, but as a secondary mechanism activated after development of large-scale convection structures.

The amplitudes of the temperature variations associated with these features can be very high. For example, even at moderately large $\Gras = 1.2\times10^8$, the flow in Fig.~\ref{fig4} shows the temperature oscillations of the amplitude about 50 K. Extrapolation to higher $\Gras$ and $\Ha$ expected in technological systems, in particular in the fusion reactors, or in astrophysical systems predicts much higher amplitudes. As a result, one should expect strong and possibly counter-intuitive effects of the magnetoconvection including new regimes of mixing and heat transfer, strong thermal stresses in the walls, changes of the rates of chemical reactions, such as corrosion, and of the phase transition behavior, activation of physical effects (e.g. thermophoresis and thermoelectric effect) typical for systems with strong temperature gradients, etc.

\section{Methods of analysis of extreme magnetoconvection}\label{sec:method}
We start with the statement that experiments in high-$\Ha$ magnetohydrodynamics are notoriously difficult. High values of $\Ha$ can only be achieved in a laboratory if liquids of high electric conductivity, i.e. liquid metals are used. Experiments are carried out with model metals, which are liquid at room (GaInSn or Hg) or slightly elevated (e.g. Na) temperatures, although high-temperature experiments, e.g. with PbLi alloy are also performed. All such liquids are opaque, so optical methods cannot be used. The commonly used tools are the ultrasonic Doppler velocimetry, potential probes, thermocouples for temperature, and the recently developed electromagnetic contactless methods: Lorentz force velocimetry \cite{ThessLFM:2007} and inductive tomography \cite{Stefani:2004}. Unfortunately, apart from the temperature measurements, the methods are either insufficiently accurate or limited in the scope of the data they generate. This often leads to the situation when an unexpected flow behavior is detected, but cannot be fully understood on the basis of the experiment alone. 

We also need to mention the rather obvious fact that the extreme magnetoconvection cannot be described in the framework of the simplified models based on the asymptotic limit (\ref{situation}). The models lead to linearized problems and inevitably predict  laminar behaviors. They allow us to describe some of the primary convection instabilities, but certainly not the non-linear development illustrated in section \ref{sec:results}.

We now consider difficulties of numerical analysis. As illustrated by the examples in section \ref{sec:results}, extreme magnetoconvection flows are often three-dimensional and unsteady, and exhibit nonlinear dynamics. This means that full three-dimensional unsteady nonlinear equations have to be solved. Simplified models, such as the quasi-two-dimensional model \cite{Sommeria:Moreau:1982} or the model of a purely two-dimensional flow in the absence of wall crossing the field lines \cite{Zhang:2015,Zhang:2017,Zhang_P1:2017,Zhang_P2:2017}, can only be used rarely, when their accuracy is verified via three-dimensional simulations or experiments. Also, since conventional three-dimensional turbulence is suppressed, no models are available for small-scale behavior, which has to be resolved.

Accurately solving the full problem at $\Ha\gg 1$, $\Gras\gg 1$ is computationally difficult, more so than in the case of non-magnetic flows or flows at moderate $\Ha$ and $\Gras$. The main reason is the numerical stiffness of the problem manifested as strong separation between the smallest and largest relevant scales. One example is the Hartmann boundary layers of thickness $\sim L\Ha^{-1}$ forming at the walls normal to the magnetic field. The recent results (see, e.g. \cite{ZikanovJFM:2013}) show that the layer must be resolved by at least 8-10 grid points for the magnetoconvection instabilities, such as, for example, illustrated in Fig.~\ref{fig2}, to be accurately reproduced. At lower resolutions, the change of the instability modes shown in Fig.~\ref{fig2} does not occur in the simulations because of incorrectly computed heat balance and temperature distribution near the wall. 
Another example is the downward flow in vertical pipes and ducts (see Fig.~\ref{fig4}). Here, good resolution of the sidewall (parallel to the imposed magnetic field) boundary layer of thickness $\sim L\Ha^{-1/2}$, which determines the structure and strength of the near-wall upward jet, is also critical. As the final comment, the illustrations in section \ref{sec:results} clearly show that the stringent numerical resolution requirements are not limited to boundary layers. Internal shear layers and mixing zones also have to be adequately resolved.

The numerical stiffness of the extreme magnetoconvection problems  means that such problems are not suitable for solution by general-purpose CFD codes (e.g. Ansys Fluent or OpenFOAM). Attempts of such simulations \cite{Di_P1:2002,Di_P2:2002,Mas:2012} have shown that convergence can only be achieved for steady-state flow regimes and at the cost of either substantial loss of accuracy or considering flows at unrealistically small $\Ha$ and $\Gras$. Extreme magnetoconvection regimes, which are typically unsteady and observed at high $\Ha$ and $\Gras$, can rarely be simulated correctly by these tools. Even when this is theoretically possible, simulations are impractical, since the low computational efficiency and slow convergence lead to excessive computational costs.

 We now turn to the case of simple geometries (cuboid, cylinder, sphere, etc). Interestingly, the problem does not appear to be amenable to the classical spectral method approach. While no systematic study of the issue has been done, it seems plausible that an accurate solution based on spectral expansions would be either unfeasible or ineffective for the  magnetoconvection flows characterized by extremely sharp gradients in boundary and internal shear layers.
 
 It has been demonstrated in the recent years that the computational difficulties can be resolved using conservative finite-volume or finite-difference schemes on structured orthogonal grids. Particularly successful has been the approach based on the finite-difference discretization principles proposed in \cite{Morinishi:1998,Ni1:2007} and systematically analyzed in \cite{Krasnov:FD:2011}. The discretization is of the second order and performed on a structured, orthogonal, non-uniform, collocated grid. Consistent approximation of differential operators is achieved via the use of fluxes of velocity and electric current density obtained by interpolation to half-integer grid points. The key to the accuracy and stability in the case of extreme magnetoconvection flows is the nearly fully conservative character of the scheme. In the non-diffusive limit, the solution exactly conserves mass, momentum, electric charge and internal energy, while kinetic energy is conserved with a dissipative error of the third order.

The numerical algorithms developed on the basis of the scheme were applied for the simulations presented in section \ref{sec:results}. Details are presented in the publications cited therein. Here, we only mention that strong clustering was used in these studies, which permitted accurate solutions on grids of reasonable size. For example, the DNS of flows in a vertical pipe shown in Fig.~\ref{fig4} were performed with the azimuthal, radial, and axial resolution by, respectively, 64, 60, and 1760 grid points.

 \section{Concluding remarks}\label{sec:conclusion}
 The recent studies of the extreme magnetoconvection reviewed in this chapter primarily concern mixed convection flows in pipes and ducts. The reasons for that are partially historical (the pioneering experiments performed for such flows in, e.g. \cite{Genin:2011Pamir,Melnikov:2016,Kirillov:2016}) and practical (the questions of the design of liquid metal blankets for fusion reactors). At the same time, the phenomenon itself is evidently much broader. As discussed in section \ref{sec:trans}, the extreme magnetoconvection behavior must be expected in every system satisfying the conditions (\ref{situation}). Many such systems are yet to be explored.
 
It would be particularly interesting to analyze the extreme natural magnetoconvection in a cylindrical or cuboid box. While active since the 1950s (see, e.g. \cite{Chandra:1961,Gershuni:1986}), studies of such systems have been largely limited to moderate values of $\Gras$ and $\Ha$ or to linearized asymptotic models, which as we have discussed above, cannot reproduce the extreme magnetoconvection behavior. A few exceptions, such as the experiments with a Raylegh-B\'{e}nard cell in vertical magnetic field \cite{Cioni:2000} have provided peculiar and yet to be explained results. We note that the presence of an imposed magnetic field and induced electric currents, as well as the technologically relevant questions mean that we need to consider several archetypal configurations. The differences between them is in the orientation of the magnetic field (vertical or horizontal), heating regime (wall heating applied at the bottom or side wall or internal heating), and the thermal and electric conductivities of the walls.

\begin{acknowledgement}
The authors are thankful to Dmitry Krasnov for the continuing support of the computational tools used for the simulations presented in this paper. The work was supported by the US NSF (Grants CBET 1232851 and 1435269) and by the Ministry of Education and Science of the Russian Federation (Project No. 13.9619.2017/8.9).
\end{acknowledgement}
%

\end{document}